\documentclass[8pt,ol,aps,pre,floatfix,twocolumn,nofootinbib]{revtex4}

\usepackage{bm}
\usepackage{epsfig}
\usepackage{graphicx} \usepackage{amsmath} \usepackage{amssymb}
\usepackage{cancel}

\def\bea{\begin{eqnarray}}
\def\eea{\end{eqnarray}}

\newcommand{\comment}[1]{}
\newcommand{\BEQ}{\begin{equation}}
\newcommand{\EEQ}{\end{equation}}
\newcommand{\BEA}{\begin{eqnarray}}
\newcommand{\EEA}{\end{eqnarray}}
\newcommand{\nn}{\nonumber \\}

\renewcommand{\d}{{\rm d}}

\newcommand{\bep}{\bar{\varepsilon}}
\newcommand{\ep}{\varepsilon}
\newcommand{\br}{{\bf r}}
\newcommand{\bn}{{\bf n}}
\newcommand{\bbm}{{\bf m}}
\newcommand{\bk}{{\bf k}}
\newcommand{\bq}{{\bf q}}
\newcommand{\bp}{{\bf p}}
\newcommand{\bs}{{\bf s}}
\newcommand{\ba}{{\bf a}}
\newcommand{\be}{{\bf e}}
\newcommand{\EE}{{\bf E}}

\newcommand{\pa}{\partial}
\newcommand{\om}{\omega}
\newcommand{\dde}{\delta^{\perp}}

\begin{document}

\title{Advantages of one and two-photon light in inverse scattering}
\author{H. Avetisyan$^{1,2}$}
\author{V. Mkrtchian$^1$}
\author{A.E. Allahverdyan$^1$\\
\textit{1. Alikahanyan National Laboratory (Yerevan Physics Institute), 2 Alikhanyan Brothers Street, \\Yerevan 0036, Armenia
}
\\\textit{2. Institute for Physical Research, Armenian National Academy of Sciences, \\Ashtarak-2, 0203, Armenia}}





\begin{abstract}
We study an inverse scattering problem in which the far-field spectral cross-correlation functions of scattered fields are used to determine the unknown dielectric susceptibility of the scattering object.
One-photon states for the incident field can resolve (at $100\%$ visibility) twice more Fourier components of the susceptibility compared to the (naive) Rayleigh estimate, provided that the measurement is performed in the back-scattering regime. Coherent states are not capable of reaching this optimal resolution (or do so with negligible visibility). Using two-photon states improves upon the one-photon resolution, but the improvement (at $100\%$ visibility) is smaller than twice, and it demands prior information on the object. This improvement can also be realized via two independent laser fields. The dependence on the prior information can be decreased (but not eliminated completely) upon using entangled states of two photons. 
\end{abstract}
\maketitle

Inverse scattering involves determining the dielectric susceptibility of the scattering object given the incident field and measuring the scattered field \cite{Devaney,baltes}. Measurements can be classified over their type (e.g. direct photodetection and intensity correlation) and distance from the object, i.e. near {\it vs.} far-field. Near-field measurements are potentially more informative than far-field ones \cite{Gilmore} (e.g. they do not hold the Rayleigh limit), but their practical implementation is more difficult and frequently demands prior information about the dielectric object. Far-field measurements allow a more universal theory that is built up around the weak-scattering limit (Born's approximation, Lippmann-Schwinger equations) \cite{Devaney}. In this regime the resolution of object's fine details for classical optical methods is governed by Rayleigh limit; e.g. experimental tomography approaches one-half of it \cite{Semenov}. 

Several methods go beyond the Rayleigh resolution limit via multi-photon light \cite{kuzmich_mandel,darya,Schotland,Abouraddy,Agarwal,Shih,Dowling,monken}. They find applications in biology and material science \cite{taylor}. In the context of interferometry (i.e.scalar, few-mode situation) using $N$-photon light increases the resolution $N$ times (e.g. via entangled N00N states) reaching Heisenberg's limit and outperforming the standard quantum limit of $\sqrt{N}$ improvement \cite{Dowling}. Entanglement is a widely studied resource for quantum-enabled optical technologies \cite{Abouraddy, takeuchi,taylor,wolfgramm,Schotland2016}, though its role might be exaggerated \cite{benni}.

We study resolution limits for a quantum field undergoing a weak scattering from an object with unknown dielectric susceptibility. Starting from the vector Helmholtz equation, we show that a one-photon state of the incident field provides two times larger resolution (at 100\% visibility) than the naive Rayleigh limit, if detected in the back-scattering regime. In this regime, the interference between the incident and scattering fields can be eliminated via polarization (with unknown dielectric susceptibility). This is an advantage of one-photon states, which is lacking for the coherent state of the incident field that cannot achieve the same optimal resolution. Next, we study two-photon states and show that they do improve the one-photon resolution. The improvement is smaller than two times, can be reproduced with phase-random coherent states, and does require prior information about the susceptibility. The amount of prior information can be reduced (but not eliminated) via entangled two-photon states. Such states do not increase the resolution {\it per se}.

{\it\bf Scattering.} A static, anisotropic, scattering object is embedded in a uniform, lossless medium. 
Helmholtz's equation for the Fourier component $E_i(\omega,\br)$ of the electric field reads \cite{Chiao}:
\BEA
\label{is1}
    \partial_k\partial_k E_i+\om^2 \ep_{ij}(\br) E_j= \partial_i\partial_k E_k, \\
    \ep_{ij}(\br)\equiv\delta_{ij}+\bep_{ij}(\br),
\EEA
where $i,k,j=1,2,3$; $\pa_i=\pa/\pa x_i$, $\bep_{ij}(\br)$ is the 
space-dependent dielectric susceptibility of the scatterer, 
repeated indices imply summation, and $c=\hbar=\epsilon_0=1$.
Eq.~(\ref{is1}) implies $\pa_i [\ep_{ij}E_j]=0$, i.e. (\ref{is1})
can be written as
\BEA
\label{is2}
&    [\partial_k\partial_k +\om^2] E_i= -\pa_i\partial_k[\bep_{kj} E_j]-\om^2 \bep_{ij}E_j, \\
&E_i(\om,\br)= -\int\d^3 r'\, G_{\om}(\br-\br') 
[ \pa_i\partial_k[\bep_{kj} E_j]\nonumber\\&
+\om^2 \bep_{ij}E_j](\br'),\\ 
&G_{\om}(\br)=-{e^{i|\om||\br|}}/(4\pi |\br|),
\label{is3}
\EEA
where $G_{\om}(\br)$ is the free out-going Green function. We separate $E_i$ into the incident and the scattered fields
\begin{align}
    {E}_i&={E}_i^{(\text{in})} + {E}_i^{(\text{s})}.
\label{etot}
\end{align}
We assume that within the dielectric the scattered field is small: ${E}_i^{(\text{in})} \gg {E}_i^{(\text{s})}$.
We then put in the RHS of (\ref{is3}): ${E}_i={E}_i^{(\text{in})}$ (first-order Born's approximation),
implement in (\ref{is3}) for $|\br||\om|\gg 1$ the far-field limit $ G_\om(\br-\br')\simeq  G_\om(\br) e^{-i|\om|\bn \br'}$, $\bn\equiv\br/|\br|$, and find 
\BEA
E_i^{\rm (s)} (\om,\br)=-G_{\om}(\br)\,\om^2\,\dde_{ik}(\bn)\int \d^3 r'\, e^{-i|\om|\bn \br'}\nonumber\\\times
\bep_{kj}(\br')E^{\rm (in)}_{j}(\om,\br'),\quad 
\dde_{ik}(\bn)\equiv\delta_{ik}-n_in_k,
\label{yalta}
\EEA
where $E^{\rm (in)}_{j}(\om,\br)$ is the free electric field operator \cite{Chiao}:
\begin{align}
\EE^{\rm (in)} (\om>0,\br)&=\int \d^3 q\, \sqrt{|\bq|} e^{i\bq\br}\, 
\be_\alpha(\bq)\, a_\alpha(\bq)\,\delta[|\bq|-\om]\nonumber \\
\label{is5}
&= |\om|^{5/2}\oint \d \Omega\, e^{i\om\br\cdot\bbm} \be_\alpha(\bbm) a_\alpha(\om\bbm),\\
\EE^{\rm (in)} (\om<0,\br)&= \EE^{\dagger {\rm (in)}} (|\om|,\br),\\
\label{door}
\be_\alpha(\bbm)\cdot \be_\beta(\bbm)&=\delta_{\alpha\beta},\quad \be_\alpha(\bbm)\cdot\bbm=0, \\
\label{is40}
[a_\alpha(\bq), a_\beta(\bq')]&=0\\
\label{is401}
[a_\alpha(\bq),
a^\dagger_\beta(\bq')]&=\delta_{\alpha\beta}\delta(\bq-\bq').
\end{align}
Eqs.~(\ref{is40},\ref{is401}) are commutation relations, 
$\oint \d \Omega$ in (\ref{is5}) means integration over directions of $|\bbm|=1$,
and where $\be_\alpha(\bq)$ in (\ref{door}) are unit polarization vectors.
Using (\ref{yalta}, \ref{is5}) we find
\begin{align}
E_i^{\rm (s)} &(\om>0,\br) =- |\om|^{9/2}G(\br)\dde_{ik}(\bn)
\label{is6}\\ 
&\times
\oint \d \Omega
{\rm e}_{\alpha|j}(\om\bbm) \bep_{kj}[\om(\bbm-\bn)] a_\alpha(\om\bbm),\\
\bep_{ij}&[\bq]\equiv\int\d^3 r\, e^{i\bq\br}\bep_{ij}(\br),\nonumber
\end{align}
which contains scatterer's Fourier transform $\bep_{kj}[\om(\bbm-\bn)]$.
The following correlations refer to photodetection via (resp.) one and two detectors located at $\br$ and $(\br_1, \br_2)$ \cite{Chiao}:
\begin{align}
\label{is7}
    &\Phi_{i}^{(1)}=
    \text{Tr}\Big[\rho E^{\dagger}_i(\omega,\textbf{r}) E_i(\omega,\textbf{r})\Big] ,\\
\label{is777}
    &\Phi_{i_1i_2}^{(2)}= \text{Tr}\Big[\rho
    E_{i_1}^{\dagger}(\omega_1,\textbf{r}_1)
    E_{i_2}^{\dagger}(\omega_2,\textbf{r}_2)
    \nonumber\\&\qquad\qquad\times
    E_{i_2}(\omega_2,\textbf{r}_2)
    E_{i_1}(\omega_1,\textbf{r}_1)\Big],
\end{align}
where $\rho$ is the initial density matrix of the field. Eqs.~(\ref{is7}, \ref{is777}) define the measured field features we focus on.

{\it\bf One-photon states} in (\ref{is7}) read 
\BEA
\label{is10}
\rho_1=|\psi\rangle\langle\psi|, ~~
|\psi\rangle=\int\d^3 q\, C_\alpha (\bq)a^\dagger_\alpha(\bq)|0\rangle,\\
\langle\psi|\psi\rangle=\int\d^3 q\, C_\alpha (\bq)C^*_\alpha (\bq)=1,
\label{is101}
\EEA
where $|0\rangle$ is the vacuum state of the field. Due to normalization 
(\ref{is101}), $|\psi\rangle$ in (\ref{is10}) is not an eigenstate of the free field Hamiltonian. Hence in the time-domain the correlators (\ref{is7}) do not hold time-translation invariance, and we do not implement the Wiener-Khinchin theorem for (\ref{is7}). Working with normalized states ensures regular transition from the original time-domain to the frequency domain (\ref{is1}). Using (\ref{is40}, \ref{is6}, \ref{is7}, \ref{is10}) we get
\begin{align}
\label{is11}
&\Phi_{i}^{(1)}= \langle \psi|  E^\dagger_i(\omega,\textbf{r})|0\rangle
\langle 0|  E_i(\omega,\textbf{r})|\psi\rangle, \\
\label{is12}
&\langle 0|  \EE^{\rm (in)}(\omega,\textbf{r})|\psi\rangle= \EE^{\rm (in)}(\omega,\textbf{r})
|_{a_\alpha(\om\bbm)\to C_\alpha(\om\bbm)},\\
&\langle 0|  E^{\rm (s)}_j(\omega,\textbf{r})|\psi\rangle= E^{\rm (s)}_j(\omega,\textbf{r})
|_{a_\alpha(\om\bbm)\to C_\alpha(\om\bbm)},
\label{is14}
\end{align}
where (\ref{is12}) and (\ref{is14}) are found from (resp.) (\ref{is5}) and (\ref{is6}).

In (\ref{is12}, \ref{is14}) we assume
$C_\alpha(\om\bbm)=c_\alpha C(\om\bbm)$, where $C(\om\bbm)$ has a sharp
maximum at $\hat\omega \bs$ with $|\bs|=1$. Hence, due to the integral $\oint$ in
(\ref{is12}) and (\ref{is14}), $\langle 0|
E_j(\omega,\textbf{r})|\psi\rangle$ is not close to zero only for
$\hat\om\simeq\om$. Given this condition, we can approximate ${\rm
e}_{\alpha|k}(\bbm) \bep_{ij}[\om(\bbm-\bn)]\simeq {\rm
e}_{\alpha|k}(\bs) \bep_{ij}[\om(\bs-\bn)]$, and take the latter
factor out of $\oint$ in (\ref{is14}):
\begin{align}
\label{gogr}
\langle 0|  E_i^{\rm (in)}(\omega,\textbf{r})|\psi\rangle &=
|\om|^{\frac{5}{2}}p_i(\bs)e^{i\om\br\cdot\bs} \oint \d \Omega\, C(\om\bbm),\\
\langle 0|  E_i^{\rm (s)}(\omega,\textbf{r})|\psi\rangle &=
- |\om|^{\frac{9}{2}}G_{\om}(\br)\dde_{ik}(\bn)p_j(\bs) \nn&\times
\bep_{kj}[\om(\bs-\bn)] \oint \d \Omega\, C(\om\bbm),
\label{gogr2}
\end{align}
where $\bp(\bs)=c_\alpha \be_\alpha(\bs)$ is the polarization
vector. The maximal resolution in (\ref{gogr2}) is achieved for 
$\bs=-\bn$. As an example, consider 
two point scattering centers located at points $\pm\ba$:
\BEA
\label{is_c}
&& \bep_{ij}[\br]=\lambda_{ij}(\delta(\br-\ba)+\delta(\br+\ba)\,), \\
&& \bep_{ij}[\om(\bs-\bn)]= 2\lambda_{ij}\cos[\om(\bs-\bn)\cdot\ba].
\label{is_cc}
\EEA
Now for $\bs=-\bn$ and (\ref{is_cc}) we get $\Phi_{i}^{(1)}\propto 1+\cos[4\om \bs\cdot\ba]$.
For $\bs\cdot\ba=\pm|\ba|$ this oscillates as $\cos[2\om (2a)]$, where $2|\ba|=2a$ is the inter-center distance. Hence the resolution is two times larger than the Rayleigh limit, i.e. the difference between the maxima and minima as a function of $\om a$ is $\frac{\pi}{4}$, while the naive Rayleigh limit gives $\frac{\pi}{2}$. Note that the visibility of this signal is $100\%$. The visibility for a signal $\mathfrak{S}(x)$ is defined as 
\BEA
[\mathfrak{S}_{\rm max}-\mathfrak{S}_{\rm min}]/[\mathfrak{S}_{\rm max}+\mathfrak{S}_{\rm min}], 
\label{visible}
\EEA
where maximization and minimization are carried over $x$. 

But $\bs=-\bn$ employed above means that $E_i^{\rm (in)}(\om,\br)$ cannot be neglected in $\Phi_{i}^{(1)}$. 
Realistically, the modes of the incident field (\ref{is5}) are not plane-waves, but rather
localized (e.g.  Gaussian) wave packets, and the widths of their
transverse profiles are large compared to the linear size of the
dielectric (hence using plane waves is legitimate), but small compared
to the distance from the dielectric to the detector located in the
far-field $\br$ \cite{shirokov}. Hence, the superposition of the incident and scattered
amplitudes (\ref{etot}) exist only in the forward $\bs\simeq\bn$ and
backward $\bs\simeq-\bn$ directions. Keeping $E_i^{\rm (in)}(\om,\br)$ makes our calculations meaningless, since now
$E_i^{\rm (s)}(\om,\br)$ can be neglected, once we work in the first Born's approximation. Moreover, if $|\langle 0| E^{\rm (in)}_j(\omega,\textbf{r})|\psi\rangle|^2$ is not negligible, $|\langle 0| E^{\rm (s)}_j(\omega,\textbf{r})|\psi\rangle|^2$ is not even a small correction to the main term $|\langle 0| E^{\rm (in)}_j(\omega,\textbf{r})|\psi\rangle|^2$, since the contribution $\langle E^{\rm (in)}_i|\psi_1\rangle\langle\psi_1| E^{\rm (s,2)}_i\rangle$ of the second-order Born term $E^{\rm (s,2)}_i$ into (\ref{is7}) has the same order of magnitude as the terms retained in (\ref{gogr2}). 

However, for one-photon states $E_i^{\rm (in)}(\om,\br)$ can be excluded due to polarization, i.e. $p_i(\om\bs)=0$ for any $\bep_{ij}[\om(\bs-\bn)]$ in (\ref{gogr}). Recalling $\bp(\om\bs)\cdot \bs=0$, we see from (\ref{gogr}) that a generic anisotropy in $\bep_{ij}$ is needed, e.g. $\bep_{ij}=\delta_{ij}\bep_i$, where $\bep_i$ depends on $i$. Otherwise, $\langle 0| E_i^{\rm (s)}(\omega,\textbf{r})|\psi\rangle=0$, and the very signal nullifies. 

We employ (\ref{is_c}, \ref{is_cc}) as our main example, but similar results are obtained also for other situations, e.g. a dielectric sphere [$\theta(x)$ is the step function]:
\BEA
\label{shar}
& \bep_{ij}[\br]=\lambda_{ij}\theta(|\br|-a)), \\
& \bep_{ij}[\bq]= {4\pi\lambda_{ij}}q^{-3}(\sin[qa]-qa\cos[qa]), ~~q=|\bq|.
\label{sharov}
\EEA

{\it\bf Coherent states} in (\ref{is7}) are defined as
\BEA
\label{c}
|\psi_{\rm c}\rangle= e^{|A|^2/2}\int\d^3 q\, f_\alpha(\bq)|A,\bq,\alpha\rangle, ~~ \langle \psi_{\rm c}|\psi_{\rm c}\rangle=1, \\
a_\alpha(\bk)|A,\bq,\alpha\rangle=A\delta(\bk-\bq)\delta_{\alpha\beta}|A,\bq,\alpha\rangle, 
\label{co}
\EEA
where $|A,\bq,\alpha\rangle=|A,\bq\rangle|\alpha\rangle$ is the coherent state with the complex amplitude $A$ in mode $\bq$, and the vacuum state over all other modes; $|\alpha\rangle$ is a pure polarization state. Now $\langle \psi_{\rm c}|\psi_{\rm c}\rangle=1$ in (\ref{c}) is due to $\langle A,\bk|A,\bq\rangle=e^{-|a|^2}$ (for $\bq\not=\bk$) and $\Big|\int\d^3 q\, f_\alpha(\bq)\Big|^2=1$.

Calculating (\ref{is7}, \ref{c}, \ref{co}) and assuming that $f(\bq)$ in (\ref{c}) concentrates at $\om \bs$ with $|\bs|=1$, we find that $\langle\psi_c|E^{{\rm (s)}\,\dagger}_i(\omega,\textbf{r}) E^{\rm (s)}_i(\omega,\textbf{r})|\psi_c\rangle$ is similar to (\ref{gogr2}). However, now for $\bn=-\bs$ (back-scattering) also the terms $\langle\psi_c|E^{{\rm (in)}\,\dagger}_i(\omega,\textbf{r}) E^{\rm (s)}_i(\omega,\textbf{r})|\psi_c\rangle$ contribute to (\ref{is7}). They cannot be neglected, since they contain the unknown $\bep_{kj}[\om(\bs-\bn)]$. Hence the maximal resolution for coherent states is not achieved via the polarization elimination of the incident field.

{\it\bf Two-photon states} are defined in (\ref{is777}) as [cf.~(\ref{is10})] 
\BEA
\label{is30}
&|\psi_2\rangle=\int C_{\alpha_1 \alpha_2}(\bq_1,\bq_2) \prod_{u=1}^2\d^3\, q_u\,
a^\dagger_{\alpha_u}(\bq_u) |0\rangle,\\
\label{west}
&C_{\alpha_1 \alpha_2}(\bq_1,\bq_2)=C_{\alpha_2 \alpha_1}(\bq_2,\bq_1),~~~ \rho_2=|\psi_2\rangle\langle\psi_2|,\\
&1=\langle\psi_2|\psi_2\rangle=2\int\d^3 q_1\d^3 q_2 |C_{\alpha_1 \alpha_2}(\bq_1,\bq_2)|^2,\\
& \Phi_{i_1i_2}^{(2)}=\Big| \left\langle 0\Big|\prod_{u=1}^2E_{i_u}(\om_u,\br_u)\Big|\psi_2\right\rangle\Big|^2,
\label{ond4}
\EEA
where (\ref{ond4}) expresses the correlation function $\Phi_{i_1i_2}^{(2)}$ via the two-photon wave-function. 
Working out (\ref{door}, \ref{is30}--\ref{ond4}) we find
\begin{align}
&\Big\langle 0\Big|\prod_{u=1}^2E^{\rm (s)}_{i_u}(\om_u,\br_u)\Big|\psi_2\Big\rangle=2\prod_{u=1}^2\om_u^{\frac{9}{2}}G_{\om_u}(\br_u)
 \nn &\quad\times
 \dde_{i_uk_u}(\bn_u)
\oint\d \Omega_u \,\bep_{k_uj_u}[\om_u(\bbm_u-\bn_u)]\nn &\quad\times
C_{\alpha_1\alpha_2}(\om_1\bbm_1, \om_2\bbm_2){\rm e}_{\alpha_1|j_1}(\bbm_1){\rm e}_{\alpha_2|j_2}(\bbm_2).~~
\label{or2}
\end{align}
We assume factorization of momenta and polarizations:
\BEA
\label{ost0}
C_{\alpha_1\alpha_2}(\om_1\bbm_1, \om_2\bbm_2)=
c_{\alpha_1\alpha_2}C(\om_1\bbm_1, \om_2\bbm_2),~~\\
c_{\alpha_1\alpha_2}=c_{\alpha_2\alpha_1}, ~
C(\om_1\bbm_1, \om_2\bbm_2)=C(\om_1\bbm_2, \om_2\bbm_1),
\label{ost}
\EEA
where (\ref{ost}) follows from (\ref{west}, \ref{ost0}). We also assume that $C(\om_1\bbm_1, \om_2\bbm_2)$ is sharply maximized at $\hat\om_1\bs_1\simeq \om_1\bbm_1$ and $\hat\om_2\bs_2\simeq \om_2\bbm_2$ or at $\hat\om_1\bs_1\simeq \om_2\bbm_2$ and $\hat\om_2\bs_2\simeq \om_1\bbm_1$, where $|\bs_1|=|\bs_2|=1$, and $\hat\om_1\not=\hat\om_2$. Then (\ref{or2}) are far from zero only for $\om_1\simeq \hat\om_1$ and $\om_2\simeq \hat\om_2$. We find from (\ref{or2}):
\begin{align}
\Big\langle &0\Big|\prod_{u=1}^2E^{\rm (s)}_{i_u}(\om_u,\br_u)\Big|\psi_2\Big\rangle
= 2\zeta[\om_1,\om_2]\prod_{u=1}^2 G_{\om_u}(\br_u) |\om_u|^{\frac{9}{2}} \nn&\qquad\qquad\times
\label{bu0}
\dde_{i_uk_u}(\bn_u)
\Theta_{j_1j_2}(\bs_1, \bs_2)\nn&\qquad\qquad\times
\bep_{k_1j_1}[\om_1(\bs_1-\bn_1)]
\bep_{k_2j_2}[\om_2(\bs_2-\bn_2)],\\
\label{bu2}
& \Theta_{i_1i_2}(\bs_1, \bs_2)\equiv c_{\alpha_1\alpha_2}
{\rm e}_{\alpha_1|i_1}(\bs_1){\rm e}_{\alpha_2|i_2}(\bs_2),\\
& \zeta[\om_1,\om_2]\equiv\oint\oint\d \Omega_1 \d \Omega_2 \, C(\om_1\bbm_1, \om_2\bbm_2),
\label{bu1}
\end{align}
where the polarization tensor $\Theta_{j_1j_2}(\bs_1, \bs_2)$ is discussed in Appendix.
For $\hat\om_1\simeq \hat\om_2\simeq \om_1\simeq \om_2=\om$, we need to take $\om_1=\om_2$ in (\ref{bu0}),
and also account for the interference terms, i.e. we should symmetrize (\ref{bu0}) and 
(\ref{bu2}) over $\bs_1$ and $\bs_2$, and divide them over 2. 

Counting in (\ref{is7}, \ref{ond4}) only the contribution (\ref{bu0}) from $E_i^{\rm (s)}$, and using the example (\ref{bu2}) with (for simplicity) isotropic condition $\lambda_{ij}=\lambda\delta_{ij}$ we find
\BEA
\label{erd} 
\Phi_{i_1i_2}^{{\rm s}\,(2)}\propto\Big| \dde_{i_1k_1}(\bn_1)\dde_{i_2k_2}(\bn_2)
\Theta_{k_1k_2}(\bs_1, \bs_2)  \Big|^2\times \\
\cos^2[a\om_1(\sigma_1-\nu_1)] \cos^2[a\om_2(\sigma_2-\nu_2)], 
\label{ferr} \\
\sigma_k=\bs_k\cdot\ba/a, ~ \nu_k=\bn_k\cdot\ba/a, ~~ a=|\ba|, ~~ k=1,2.
\label{ipso}
\EEA
\begin{figure}
\begin{center}
\includegraphics[width=0.6\linewidth]{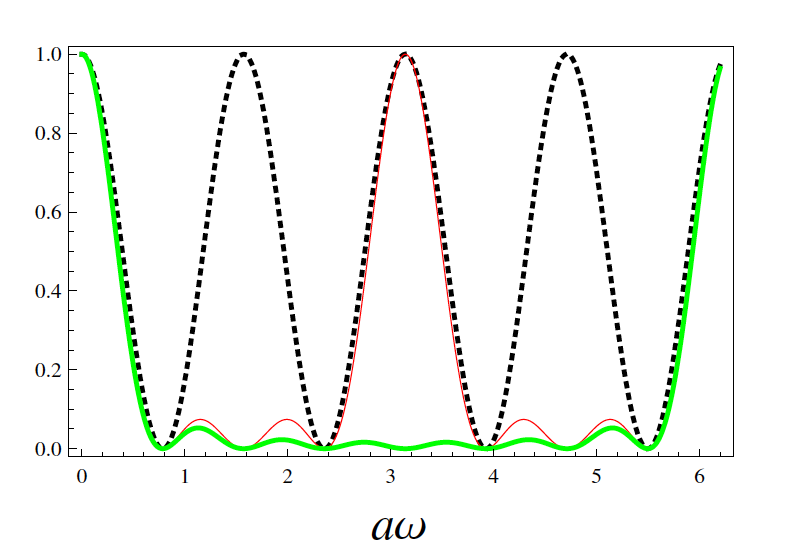}
\caption{Resolution regimes. Red (thin) curve: $\cos^2[2a\om\chi]\cos^2[a\om\chi]$, where $a=|\ba|$ and $\chi=0.9$; cf.~(\ref{bu2}, \ref{hall}). Black (dotted) curve: $\cos^2[2a\om\chi]$. 
Green (solid) curve: $\cos^2[2a\om\chi]\cos^2[a\om\chi]\cos^2[a\om\chi/2]$; cf.~(\ref{hev}).
}
\hfill
\label{fig1}
\end{center}
\end{figure}

When looking at the resolution provided by (\ref{ferr}), we should recall that $a$ is an 
unknown parameter that is to be estimated by looking at the difference between maxima and
minima of (\ref{ferr}). The sub-optimal resolution in (\ref{ferr}) that is achieved for a possibly large
domain of $a$ reads: 
\BEA
\sigma_1-\nu_1=2\chi, \quad \om_2(\sigma_2-\nu_2)=\om_1 \chi, \quad \frac{1}{2}< |\chi|< 1,
\label{hall}
\EEA
where it is natural to assume $\om_1>\om_2$, and then (\ref{hall}) implies $\om_2>\om_1/2$. Note that for increasing the resolution it is desirable to take $|\chi|=1$, but this cannot be done, since $E_i^{\rm (in)}$ cannot be neglected in (\ref{ond4}). Now we need to eliminate three quantities:
\BEA
\Big\langle 0\Big|E^{\rm (u)}_{i_1}(\om_1,\br_1)E^{\rm (v)}_{i_2}(\om_2,\br_2)\Big|\psi_2\Big\rangle,
\\ 
({\rm u}, {\rm v})=({\rm in}, {\rm in}), ({\rm s}, {\rm in}), ({\rm s}, {\rm s}).\nonumber
\nonumber
\EEA
These three quantities cannot be eliminated simultaneously via polarization degrees of freedom for a generic $\bep_{ij}$. Hence, in contrast to the one-photon situation, we have to eliminate $E_i^{\rm (in)}$ via avoiding detection in forward and backward directions, i.e. avoiding $\sigma_1=\pm \nu_1$ and $\sigma_2=\pm \nu_2$. This implies $|\chi|< 1$ in (\ref{hall}). Now for $a\om\in\mathfrak{D}_{\mathfrak{n}}\equiv \frac{1}{\chi}[\frac{\pi}{4}+\pi \mathfrak{n}, \frac{3\pi}{4}+\pi \mathfrak{n}]$, where $\mathfrak{n}$ is an integer, (\ref{ferr}) predicts resolution $\frac{\pi}{8\chi}$; see Fig.~\ref{fig1}. This is the difference between maxima and minima of (\ref{ferr}) as a function of $a\om\in \mathfrak{D}_{\mathfrak{n}}$. For $\chi>\frac{1}{2}$ [cf.~(\ref{hall})] this is better (smaller) than the one-photon resolution $\frac{\pi}{4}$; see after (\ref{is_cc}). For  $a\om\not\in\mathfrak{D}_{\mathfrak{n}}$, the resolution $\frac{\pi}{4\chi}$ predicted by (\ref{ferr}) is worse (larger) than the one-photon resolution $\frac{\pi}{4}$. Thus for increasing the resolution via the two-photon state, we need prior information $a\om\in\mathfrak{D}_{\mathfrak{n}}$. 

The visibility (\ref{visible}) for (\ref{ferr}) is still $100\%$. With two-photon states, we did not find any better resolution than $\frac{\pi}{8\chi}$, even when the interference terms (for $\om_1=\om_2$) were involved in (\ref{bu0}, \ref{bu2}). We emphasize that the above comparison with one-photon situation using $\om_1>\om_2$ is a fair one, because the frequency is a resource for resolution. 

The above improvement in resolution (with 100 \% visibility) can also be obtained via independent coherent states of the field, which is the set-up of the Pfleegor-Mandel experiment \cite{pfleegor-mandel-b,ou}. Here for the initial state of the field we have [cf.~(\ref{c}, \ref{co})]: 
\begin{align}
|\psi_{2{\rm c}}\rangle&= e^{|A|^2}\prod_{u=1}^2c^{(u)}_\alpha\int\d^3 q_u\, \nn&\qquad\times 
f^{(u)}(\bq_u)|A_1,\bq_1,\alpha_1; A_2,\bq_2,\alpha_2 \rangle,
\label{coco}
\end{align}
where the latter two-mode coherent state is defined by analogy to (\ref{co}), $|A_1|^2=|A_2|^2=|A|^2$ (for simplicity), and where we already assumed the factorization for polarizations: $f^{(u)}_\alpha(\bq_u)=c^{(u)}_\alpha f^{(u)}(\bq_u)$. Formulas similar to (\ref{ferr}) are found from (\ref{coco}) upon assuming that $f^{(u)}(\bq_u)$ concentrates at $\om_u\bs_u$ for $u=1,2$ and $\om_1\not=\om_2$. For this result, we can also assume  $A_1=|A|e^{i\phi_1}$ and $A_2=|A|e^{i\phi_2}$ with random phases $\phi_1$ and $\phi_2$. The known result \cite{ou} on the visibility decrease for the classical light does not apply to our situation, since we do not consider interference.

{\it\bf Entangled biphotons.} Now consider the case when (\ref{or2}--\ref{ost}) contains a superposition of at least two biphotons. (We did not focus on the superposition of one-photon states,
since it does not lead to any resolution improvement.) In (\ref{or2}--\ref{ost}) we assume:
\BEA
&C(\om_1\bbm_1, \om_2\bbm_2)=R(\om_1(\bbm_1-\bq_1))R(\om_2(\bbm_2-\bq_2))\nn
&+R(\om_1(\bbm_1-\bs_1))R(\om_2(\bbm_2-\bs_2)), \quad \om_1\not=\om_2,
\label{klin}
\EEA
where $R({\bf k})=R(|{\bf k}|)$ is concentrated around $|{\bf k}|=0$ and we omitted the 
symmetric part of $C(\om_1\bbm_1, \om_2\bbm_2)$ in (\ref{klin}), since for 
$\om_1\not=\om_2$ it does not contribute into the integral $\oint$ in (\ref{or2}).
Similar to (\ref{is_c}--\ref{ferr}) we obtain [cf.~(\ref{ipso})]
\begin{align}
&\Phi_{i_1i_2}^{{\rm s}\,(2)}\propto\Big| \dde_{i_1k_1}(\bn_1)\dde_{i_2k_2}(\bn_2)
\Theta_{k_1k_2}(\bq_1, \bq_2)
\nn&\quad\times
\Big(
\cos[a\om_1(\kappa_1-\nu_1)]
\cos[a\om_2(\kappa_2-\nu_2)] \nn&\quad+ 
\cos[a\om_1(\sigma_1-\nu_1)] 
\cos[a\om_2(\sigma_2-\nu_2)] \Big)
\Big|^2,\label{kamo} 
\nn& 
\kappa_1=(\bq_1\cdot\ba)/a, \quad \kappa_2=(\bq_2\cdot\ba)/a,
\end{align}
where we took $\Theta_{k_1k_2}(\bq_1, \bq_2)=\Theta_{k_1k_2}(\bs_1, \bs_2)$ 
[checked this with (\ref{sir})]. Assuming in (\ref{kamo}) $\om_1>\om_2$, 
(\ref{otto}) and (\ref{michelen}), we find (\ref{hev}):
\BEA
\label{otto}
&\kappa_1=\sigma_1, \quad \kappa_1-\nu_1=2\chi, \quad \frac{1}{2}< |\chi|< 1,\\
&\om_2(\sigma_2-\kappa_2)=\om_1\chi, \,\, \om_2(\kappa_2+\sigma_2-2\nu_2)=2\om_1\chi,
\label{michelen}\\
& \Phi_{i_1i_2}^{{\rm s}\,(2)}\propto \cos^2[2a\om_1\chi]\cos^2[a\om_1\chi] \cos^2\left[\frac{a}{2}\om_1\chi\right].
\label{hev}
\EEA
Fig.~\ref{fig1} shows how (\ref{hev}) behaves as a function of $a\om$. It improves upon (\ref{hall}, \ref{ferr})
in the following sense: the resolution (at 100\% visibility) provided by (\ref{hev}) is still $\frac{\pi}{8\chi}$, as for non-entangled photons. But the region, where his resolution is achieved is now larger, i.e. we need less prior information. 

{\it\bf Conclusion.} 
We aimed to understand how much the state of the quantum field can improve the resolution in inverse optics. Here spectral correlation measurements of the weakly scattered field (in the far-field limit) are employed for determining the dielectric susceptibility of the scatterer. Our analysis shows that the single photon state has an advantage here. It beats the naive Rayleigh limit, while the maximal (twice) resolution improvement is obtained provided that the photodetection of the scattered wave is done in the back-scattering regime and the dielectric susceptibility is generically anisotropic. In this regime the interference with the incident light seems inevitable. If present, this interference will ruin the information provided by the scattered wave by diminishing its visibility. But for one photon-states (and basically only for them) the interference can be excluded due to polarization. A deeper understanding of these issues demands paraxial quantization, which is not attempted here. Using two-photon states (also semi-classical states coming from two independent lasers) we are able to improve the resolution less than two times (over the one-photon state), and only for certain ranges of parameters, i.e. when prior information is available. One-photon entanglement appeared to be useless for improving resolution, while two-photon entanglement reduces the amount of prior information, but does not improve the resolution {\it per se}. Also effects related to photon indistinguishability did not improve the resolution. 

{\it\bf Appendix: Polarization tensor}
$\Theta_{j_1j_2}(\bs_1, \bs_2)$ in (\ref{bu2}) simplifies if we take $\bs_1=(0,0,1)$, 
locate $\bs_2$ in the $x_3x_2$-plane, and choose:
\BEA
\label{brut}
\be_1(\bs_1)= \be_1(\bs_2)=(1,0,0), \qquad \be_2(\bs_1)=(0,1,0), \\
\bs_2=(0,\cos\varphi,\sin\varphi), ~~~
\be_2(\bs_2)=(0,\cos\varphi,-\sin\varphi), ~
\label{brutus}
\EEA
where $\varphi$ is a parameter. We find from (\ref{bu2}, \ref{ost}, \ref{brut}, \ref{brutus}):
\BEA
\Theta_{ij}(\bs_1, \bs_2)=(c_{11}\delta_{i1}+c_{12}\delta_{i2})\delta_{j1}\nn
+(c_{22}\delta_{i2}+c_{12}\delta_{i1})(\delta_{j2}\cos\varphi
-\delta_{j3}\sin\varphi). 
\label{sir}
\EEA

Funding:
This work was supported by SCS of Armenia, grants No. 20TTAT-QTa003 and No. 21AG-1C038, and Faculty Research Funding Program 2022 implemented by the Enterprise Incubator Foundation with the support of PMI Science.

Acknowledgments:
We thank D. Petrosyan and M.Rafayelyan for discussions. 


\smallskip

\bibliography{sample}

\end{document}